\documentclass[11pt]{article}
\usepackage{psfig}
\newcommand{\final}{%           Standard double spacing in thesis
        \renewcommand{\baselinestretch}{1.5}%
        \small\normalsize%
}
\begin{document}
\title{\bf  Origin of PSRs with 0.1 $<$ P $<$ 0.3 s and 5$\times$10$^5$ $<$ $\tau$ $<$ 10$^7$ yr after
the second Supernova explosion in HMXBs.
} \author{O.H. Guseinov,
$\sp{1,2}$ \thanks{e-mail:huseyin@gursey.gov.tr}
A.O.Allakhverdiev,
$\sp{3}$ \thanks{e-mail:physic@lan.ab.az} \\
%Efe Yazgan$\sp2$
%\thanks{e-mail:yazgan@astroa.physics.metu.edu.tr}
Sevin\c{c} Tagieva$\sp3$
\thanks{email:physic@lan.ab.az}
%astro@physics.ab.az}
, \"{O}zg\"{u}r Ta\c{s}k\i n$\sp4$
\thanks{email:ozgur@astroa.physics.metu.edu.tr} \\ \\
{$\sp1$T\"{U}B\.{I}TAK Feza G\"{u}rsey Institute,} \\
{81220,\c{C}engelk\"{o}y,\.{I}stanbul,Turkey} \\ \\
{$\sp2$Akdeniz University, Department of Physics,} \\
{Antalya, Turkey} \\
{$\sp3$Academy of Science, Physics Institute,} \\
{Baku 370143, Azerbaijan Republic} \\ \\
{$\sp4$Middle East Technical University, Department of Physics,} \\
{Ankara 06531, Turkey} \\
}

\maketitle
\final
\newpage
\begin{abstract}
\noindent The origin of a part of the single pulsars with
relatively low magnetic fields $B< 10^{12} G$ and with
characteristic times $\tau <10^7$ yr is established. Such pulsars
occur as a result of the disruption of High Mass X-ray Binary Systems
after a second Supernova explosion. In these binaries mass accretion
onto the surface of X-ray pulsars leads to the decrease of the
magnetic field from its initial value $B\sim10^{12} -10^{13}$ G down
to $B< 10^{12}$ G similar to the processes in Low Mass X-ray Binaries.
\end{abstract}

%\clearpage
\parindent=0.2in
\section{Introduction}
As known the first High Mass X-ray Binary (HMXB), Scorpion X-1,
was discovered in 1962 (Giacconi et al., 1963). Large number of HMXBs was discovered
with rockets before the launch of the first X-ray satellite UHURU
in 1971. This is understandable since the angular resolution of
the rocket observations being low make the uncertainties in the
coordinates of the X-ray sources as high as 5 degrees and massive
stars can easily be observed for identification in optic band.
After UHURU started to operate many X-ray sources with low fluxes
were also discovered. Their origin is considered to be
extragalactic. The population of Low Mass X-ray Binaries (LMXBs),
their Nova type explosions, and also the light curves of them were
predicted (Amnuel et al., 1973; Amnuel et al., 1974) from the data 
analysis of the sources which were
detected by UHURU (Giacconi et al., 1973) and also using the poor data of Nova Cen
X-2 (Harriees 1967; Rao et al., 1969)  and of Cen X-4 (Evans et al., 1970). 
Investigations of LMXBs led to
very important results not only for X-ray astronomy, but also for
understanding the population of radio pulsars and the evolution of
their magnetic field. Large characteristic times for the magnetic
field decay and the alignment of magnetic and rotation axes were
understood. Bisnovatyi-Kogan \& Komberg (1974 and 1976) showed that
accretion of matter in LMXBs must lead to the strong decay of
magnetic field of neutron stars. They also predicted the spinning
up of the neutron stars in LMXBs and the existence of millisecond
radio pulsars (PSRs) with magnetic field B$<$3$\times$10$^{10}$ G.

During the last $\sim$ 40 years the evolution of single PSRs has
been studied on the spin period versus time derivative of the spin
period (P-\.{P}) diagram. Today we may roughly explain the
locations of PSRs on different parts of the P-\.{P} diagram.
However, it is still difficult to understand why a large number of
PSRs have periods in the interval 0.1-0.3 sec and magnetic field
between 10$^{11}$ G and 10$^{12}$ G. We assert that part of these
PSRs may appear after a second Supernova explosion in HMXBs. After
the explosion of the massive component, X-ray pulsar must turn
into a radio pulsar.

\section{Testing the origin of the pulsars with magnetic fields between
10$^{11}$-10$^{12}$ G and periods between 0.1-0.3 sec.}

A P-\.{P} diagram is plotted in Figure 1 for the PSRs which
have period derivatives between 10$^{-13}$ s/s and 10$^{-16}$ s/s
and distances up to 5 kpc. These restrictions were chosen in order
to deal only with a sample of single pulsars and to see clearly
the regions where the examined PSRs are located. We had to put a
limit on distance values since the farther the PSRs are the larger
the errors in their distances are. This was necessary if we also
take into consideration that in the recent surveys many distant
pulsars were discovered in the plane of the Galaxy in a narrow
latitude interval (for the complete data see ATNF Pulsar Catalogue 2003; 
Guseinov et al., 2002 and the references therein). 
There is no search with similar sensitivity for most of the parts of the sky.

As it is seen in Figure 1, all the PSRs with ages up to $10^5$
years have magnetic fields larger than $10^{12}$ G. From this
figure we can say that PSRs are practically always born with such
magnetic field values. All 23 PSRs which have genetic connections
with Supernova remnants (Manchester et al., 2001; Guseinov et al., 2003) 
are located on this part of the P-\.{P} diagram. 
Therefore, without any doubt PSRs are mostly born in this region of 
the P-\.{P} diagram. But this does not mean
that right after a Supernova explosion some PSRs can not have
several times smaller magnetic field values at birth.

As seen in Figure 1, the number of pulsars which are located in
between the constant magnetic field lines $10^{11}-10^{12}$ G and
have ages up to 5$\times$10$^5$ yr is very small, whereas the
number of pulsars having the same magnetic field values with
larger ages is high. In contrast we do not see a similar situation
in $10^{12}<B<10^{13} G$ interval. As magnetic fields of single
PSRs with ages 5$\times$10$^5$-10$^7$ yr practically do not
change, their evolutionary tracks must be parallel to the constant
magnetic field lines. Therefore, part of the PSRs with $B\simeq
10^{11}-10^{12} G$ may be born in the region with $P\simeq 0.1-0.3
sec$ directly with large value of $\tau$. This may be due to the
second Supernova explosion in HMXBs. In this way, after the
explosion in the massive binary, X-ray pulsar becomes a radio
pulsar. If we compare a pulsar having such origin with a single
born pulsar, the former must have several times smaller magnetic
field compared to the single born pulsar even though both of them
have the same initial magnetic field strengths, because there
occurs magnetic field decay due to the accretion during the HMXB
phase. This process is similar to the one that takes place during
disk accretion in LMXBs (Bisnovatyi-Kogan and Komberg 1974; 
Bisnovatyi-Kogan and Komberg 1976). 

Kinematic ages of these PSRs on average must be smaller than the
kinematic ages of the single born pulsars with the same values of
characteristic times ($\tau$). This is reasonable since the space
velocities of the single born pulsars are considerably larger than
the center of mass velocity of HMXBs. We should also take into
account that the radio pulsar which is born after the second
Supernova explosion continues to move with its orbital velocity,
but on average the orbital velocities are also smaller than the
space velocities of single born PSRs. Therefore, PSRs which appear
after the second Supernova explosion may have on the average
several times smaller distances from the Galactic plane
($\mid$z$\mid$).

In order to test this idea we have chosen two groups of PSRs with
similar characteristic ages. The number of PSRs in each group is
almost the same. In the P-\.{P} diagram the boundary of the
first group is determined as $0.1<P<0.3$ sec, $5\times 10^5 <\tau
<10^7$ yr and the second group has the boundaries $0.6<P<1.3$ sec,
$6\times 10^5 <\tau <10^7$ yr and $\dot{P}>3 \times 10^{-15}$ s/s.
It is necessary to note that we specially restricted the groups we
are working on, so that, the pulsars in the first and the second
groups have different bands of evolutionary tracks on the
P-\.{P} diagram and their number versus age distributions are
similar.

For our pulsar samples the $\mid$z$\mid$-$\tau$ diagrams are represented in
Figures 2a and 2b. As it is seen, there is no considerable
increase in $\mid$z$\mid$ with increasing $\tau$ for the first group of
PSRs. At a smaller value of $\tau$, $\mid$z$\mid$ increases approximately
with $\tau$, but there is no linear proportionality similar to the
one seen in the second group of PSRs (Figure 2b). As it is seen in
Figure 2a, at every value of $\tau$ there is a large number of
PSRs with small $\mid$z$\mid$ values. Due to their high velocities the
part of single born PSRs which are near the Galactic plane must
decrease as we go towards higher $\tau$ values (see Figure 2b).
From Figure 2a we noticed that PSRs in the first group have
definitely smaller average $\mid$z$\mid$ values compared to the PSRs of
the second group (see Figure 2b). This should be related to the
appearance of newborn PSRs in the first group with different
values of $\tau$. These PSRs are directly placed in the
low-$B$-large-$\tau$ part of the P-\.{P} diagram.

For the sake of safety we checked PSRs with $10^7<\tau <10^8$
years and saw the same tendency. However, at such ages the changes
in the beaming factor and the decrease in the pulsar voltage start
to affect the pulsar population creating uncertainties. Naturally,
this may have influence on the average value of the deviation from
the Galactic plane.

Do the important differences that we saw in Figure 2a and in
Figure 2b between the PSRs in the first and in the second groups
depend on selection effects? Can it be true that different
velocities and average $\mid$z$\mid$ values are due to other parameters of
these PSRs being different? Even though this idea has a low
probability of occurrence let us investigate other parameters of
PSRs in both groups in order not to leave any suspects. Can the
locations of the PSRs in the first group with respect to the Sun
be responsible for their small $\mid$z$\mid$ values when compared to the
$\mid$z$\mid$ values of the PSRs in the second group? In Figure 3a and in
Figure 3b the distance from the Sun (d) versus the galactic
longitude (l) of these two samples are plotted. From these figures
we deduce that PSRs in the second group are a little bit farther
away from the Sun than the PSRs in the first group. Thus, the
distances have no effect on our results.

In general it is known that radio luminosities of PSRs depend very
weakly on their locations on the P-\.{P} diagram. In Figure 1, the
locations of the PSRs in both of the groups slightly, but not
significantly, differ. Therefore, PSRs in both groups must have
similar fluxes and luminosities. On the other hand, conditions for
PSR observation depend on the directions in the Galaxy. This is
expected since in different directions of the Galaxy pulsar
searches with varying accuracies and sensitivities were performed
at different frequencies. Luminosities at 400 MHz versus Galactic
longitude are shown in Figure 4a and in Figure 4b whereas
luminosities at 1400 MHz versus Galactic longitude are represented
in Figure 5a and in Figure 5b for our samples. As seen from these
figures there is no difference for both groups in their
directions, luminosities and fluxes. Therefore, the differences in
the average values of $\mid$z$\mid$ for both groups of PSRs are not
related to the selection effects.

\section{The difference in space velocities of the pulsars in the groups}

Now let us discuss data about the space velocities of some PSRs
from both of the groups. Some data for 5 PSRs from the first group
and for 12 PSRs from the second group for which proper motions are
known are given in Table 1. As the table indicates PSRs from the
first group not only have considerably small $\mid$z$\mid$ values but also
have small space velocities. The space velocity of each pulsar may
roughly be estimated using the proper motion (Harrison et al., 1993;
Fomalont et al., 1997; Lyne et al., 1982; Bailes et al., 1990) and
distance values (Guseinov et al., 2002). Therefore, we saw that kinematic
characteristics of these two groups of pulsars are actually
different.

\section{Examination of the X-ray data of the pulsars in both groups}

After the Supernova explosion, as the X-ray component in HMXB may
directly show itself in the regions where $\tau$ values are high,
it may have significantly large X-ray luminosity compared to other
pulsars with similar characteristic times. Because the time it has
spent as a radio pulsar is smaller than the value of its
characteristic time.

Lists of PSRs from which X-ray radiation was observed in 0.1-2.4
keV and in 2-10 keV bands were published by Becker and Trumper
(1997), Becker and Aschenbach (2002) and Possenti et al., (2002).
As seen in these lists, from PSRs J1057-5226, J0358+
5413, J0538+2817 and J1932+1059 X-ray radiation was detected. All
of these PSRs belong to our first group and there is no other
pulsar with $\tau$ in the interval $5\times 10^5 -10^7 yr$ from
which X-ray radiation was observed. In the mentioned lists there
are yet two PSRs, namely J0826+2637 and J0953+0755, with several
times larger characteristic times. Both of these pulsars are
directly located in the belt along which PSRs from the first
sample evolve. Yet 3 PSRs, namely J1952+3252, J0117+5914 and
J1302-6350, which radiate X-rays, are located on the P-\.{P}
diagram right in front of the first sample of pulsars, as they
have 10$^5$$<$$\tau$$<$5$\times$10$^5$ yr. Only Geminga pulsar
which has a magnetic field about 1.7$\times$10$^{12}$ G is located
in the boundary of the second sample of PSRs. All other PSRs with
larger ages from which X-ray radiation have been observed are old
millisecond pulsars. These deductions strongly justify what we
have suggested about the origin of a great deal of the PSRs that
are born with magnetic fields between $10^{11}-10^{12}$ G and
characteristic ages between $5\times 10^5 -10^7$ years.

\section{Conclusions}

Part of the PSRs with $10^{11}<B<10^{12}$ G and with $P<0.3$ sec
appears after the second Supernova explosion in HMXBs. These PSRs,
before the second explosion, were X-ray components of HMXBs. As a
result of accretion onto these pulsars, there occurs magnetic
field decay. These PSRs must conserve their orbital and
center-of-mass velocities, the sum of which is on average smaller
than the space velocity of single born pulsars. On the other hand,
distances of HMXBs from the plane of Galaxy are not large and real
ages of the discussed pulsars may be smaller than their
characteristic times. Therefore, these PSRs must have small space
velocities and their distances from Galactic plane must also be
smaller than the single born pulsars with the same values of
$\tau$. On the other hand, they must have significantly larger
X-ray luminosity than the single born PSRs with similar values of
$\tau$. The observational data confirm all these expectations.

\clearpage
\begin{table}
%\rotate{
\begin{tabular}{lcllc}\hline\hline
\multicolumn{1}{l}{\ PSR} & \multicolumn{1}{c}{\ $\mu$} &
\multicolumn{1}{l}{\ d} & \multicolumn{1}{l}{\ $\mid$z$\mid$} &
\multicolumn{1}{l}{\ log $\tau$} \\ \hline & (mas $yr^{-1}$) &
(kyr) & (kyr) & \\ \hline
J0358+5413 & 13.9 (1) & 2 & 0.028 & 5.75 \\
J1453-6413 & 26.9 (2) & 1.84 & 0.142 & 6.01 \\
J1559-4438 & 14.0 (1) & 1.63 & 0.181 & 6.60 \\
J1932+1059 & 88.1 (3) & 0.2 & 0.013 & 6.49 \\
J2055+3630 & 4.2 (4) & 4.2 & 0.409 & 6.98 \\
& & & \\
& & & \\
J0502+4654 & 11.3 (4) & 1.7 & 0.091 & 6.26 \\
J0630-2834 & 38.1 (1) & 1.8 & 0.519 & 6.44 \\
J0653+8051 & 19.0 (4) & 2.3 & 1.038 & 6.70 \\
J0837+0610 & 51.0 (3) & 0.6 & 0.266 & 6.47 \\
J0946+0951 & 43.4 (3) & 0.98 & 0.670 & 6.69 \\
J1136+1551 & 371.3 (3) & 0.24 & 0.224 & 6.70 \\
J1509+5531 & 99.8 (3) & 1.4 & 1.108 & 6.37 \\
J1709-1640 & 3.0 (1) & 0.9 & 0.212 & 6.21 \\
J1913-0440 & 8.6 (4) & 3.1 & 0.384 & 6.51 \\
J1919+0021 & 2.2 (4) & 2.95 & 0.316 & 6.42 \\
J2225+6535 & 182.4 (4) & 2.0 & 0.238 & 6.05 \\
J2354+6155 & 22.8 (4) & 3.1 & 0.010 & 5.96 \\ \hline
\end{tabular}
%}

\vspace{0.5cm} {Table 1. Name, proper motion, d, $\mid$z$\mid$ and
$\tau$ values of 5 pulsars from the first sample and of the
following 12 pulsars from the second sample. The numbers in
parenthesis in the second column indicate the references for the
proper motions.}

{(1) Fomalont et al., (1997); (2) Bailes et al., (1990); (3) Lyne et al., (1982);
(4) Harrison et al., (1993) \\}
\end{table}

\clearpage {\bf Figure Captions}

\noindent {\bf Figure 1}: The $P-\dot{P}$ diagram for all pulsars
that have distances up to 5 kpc with
10$^{-16}$$<$\.{P}$<$10$^{-13}$ s/s. Pulsars that form our first
sample are shown with '+' sign in the figure. They have
$0.1<P<0.3$ sec, 5$\times$10$^5$ $<$ $\tau$ $<$ 10$^7$ yr. The
second sample that we consider appear with a '$\circ$' sign. These
pulsars have $0.6<P<1.3$ sec, $\dot{P}$ $>$ 3$\times$10$^{-15}$
s/s, 6$\times$10$^5$ $<$ $\tau$ $<$ 10$^7$ yr.

\noindent {\bf Figure 2a}: $\mid$z$\mid$-$\tau$ diagram for the pulsars
with $0.1<P<0.3$ sec, 5$\times$10$^5$ $<$ $\tau$ $<$ 10$^7$ yr and
$d\leq 5$ kpc.

\noindent {\bf Figure 2b}: $\mid$z$\mid$-$\tau$ diagram for the pulsars
with $0.6<P<1.3$ sec, $\dot{P}$ $>$ 3$\times$10$^{-15}$ s/s,
6$\times$10$^5$ $<$ $\tau$ $<$ 10$^7$ yr and $d\leq 5$ kpc.

\noindent {\bf Figure 3a}: The d versus l diagram for the pulsars
with $0.1<P<0.3$ sec, 5$\times$10$^5$ $<$ $\tau$ $<$ 10$^7$ yr and
$d\leq 5$ kpc.

\noindent {\bf Figure 3b}: The d versus l diagram for the pulsars
with $0.6<P<1.3$ sec, $\dot{P}$ $>$ 3$\times$10$^{-15}$ s/s,
6$\times$10$^5$ $<$ $\tau$ $<$ 10$^7$ yr and $d\leq 5$ kpc.

\noindent {\bf Figure 4a}: The luminosity at 400 MHz $L_{400}$
versus l diagram for the pulsars with $0.1<P<0.3$ sec,
5$\times$10$^5$ $<$ $\tau$ $<$ 10$^7$ yr and $d\leq 5$ kpc.

\noindent {\bf Figure 4b}: The luminosity at 400 MHz $L_{400}$
versus l diagram for the pulsars with $0.6<P<1.3$ sec, $\dot{P}$
$>$ 3$\times$10$^{-15}$ s/s, 6$\times$10$^5$ $<$ $\tau$ $<$ 10$^7$
yr and $d\leq 5$ kpc.

\noindent {\bf Figure 5a}: The luminosity at 1400 MHz $L_{1400}$
versus the galactic longitude l diagram for the first sample in
which the pulsars have $0.1<P<0.3$ sec, 5$\times$10$^5$ $<$ $\tau$
$<$ 10$^7$ yr and $d\leq 5$ kpc.

\noindent {\bf Figure 5b}: The luminosity at 1400 MHz $L_{1400}$
versus l diagram for the pulsars with $0.6<P<1.3$ sec, $\dot{P}$
$>$ 3$\times$10$^{-15}$ s/s, 6$\times$10$^5$ $<$ $\tau$ $<$ 10$^7$
yr and $d\leq 5$ kpc.

\clearpage
\begin{figure}
%\begin{tabular}{l}
\psfig{file=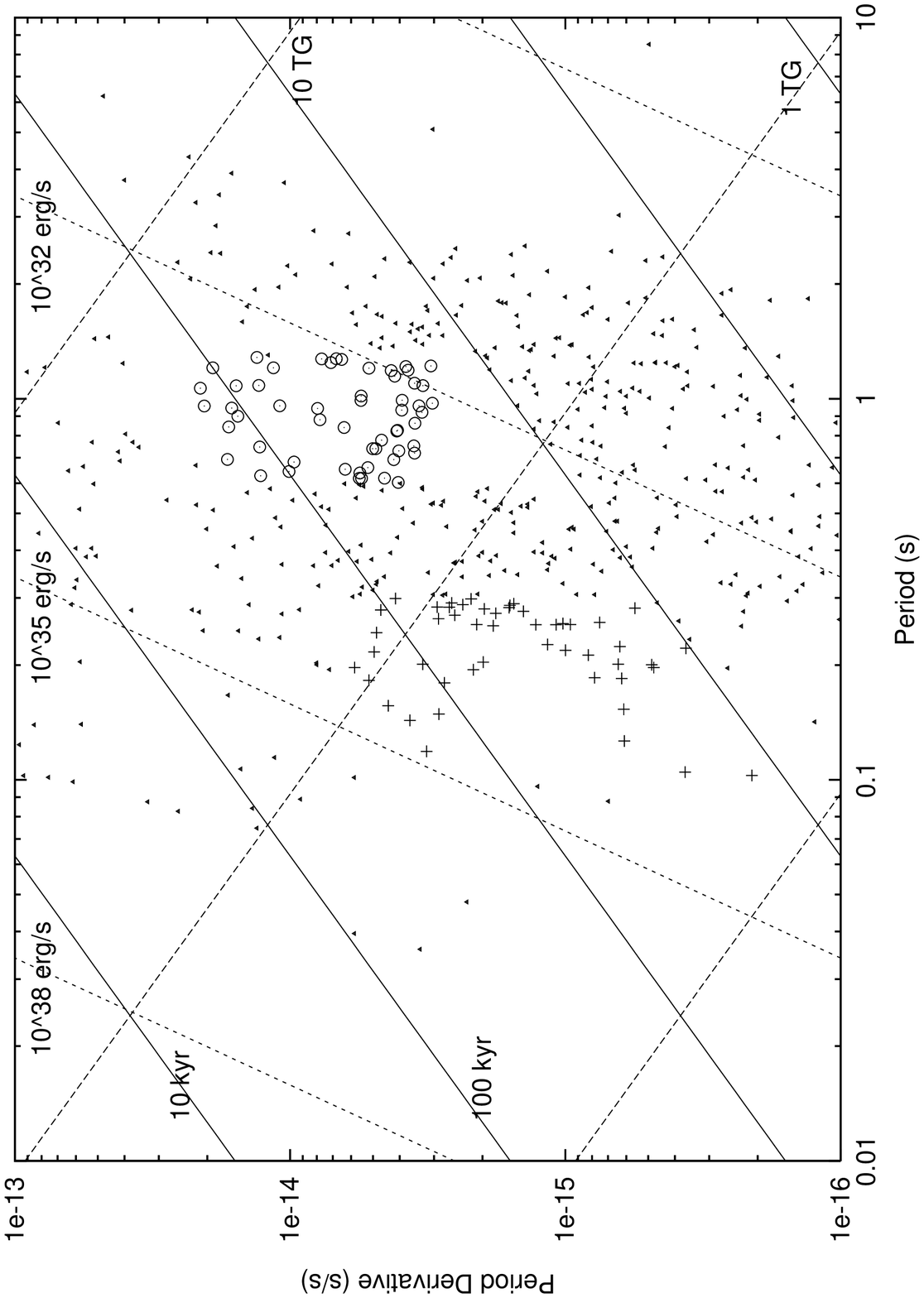,width=14.5cm,height=19cm,angle=0}
\end{figure}

\clearpage
\begin{figure}
\psfig{file=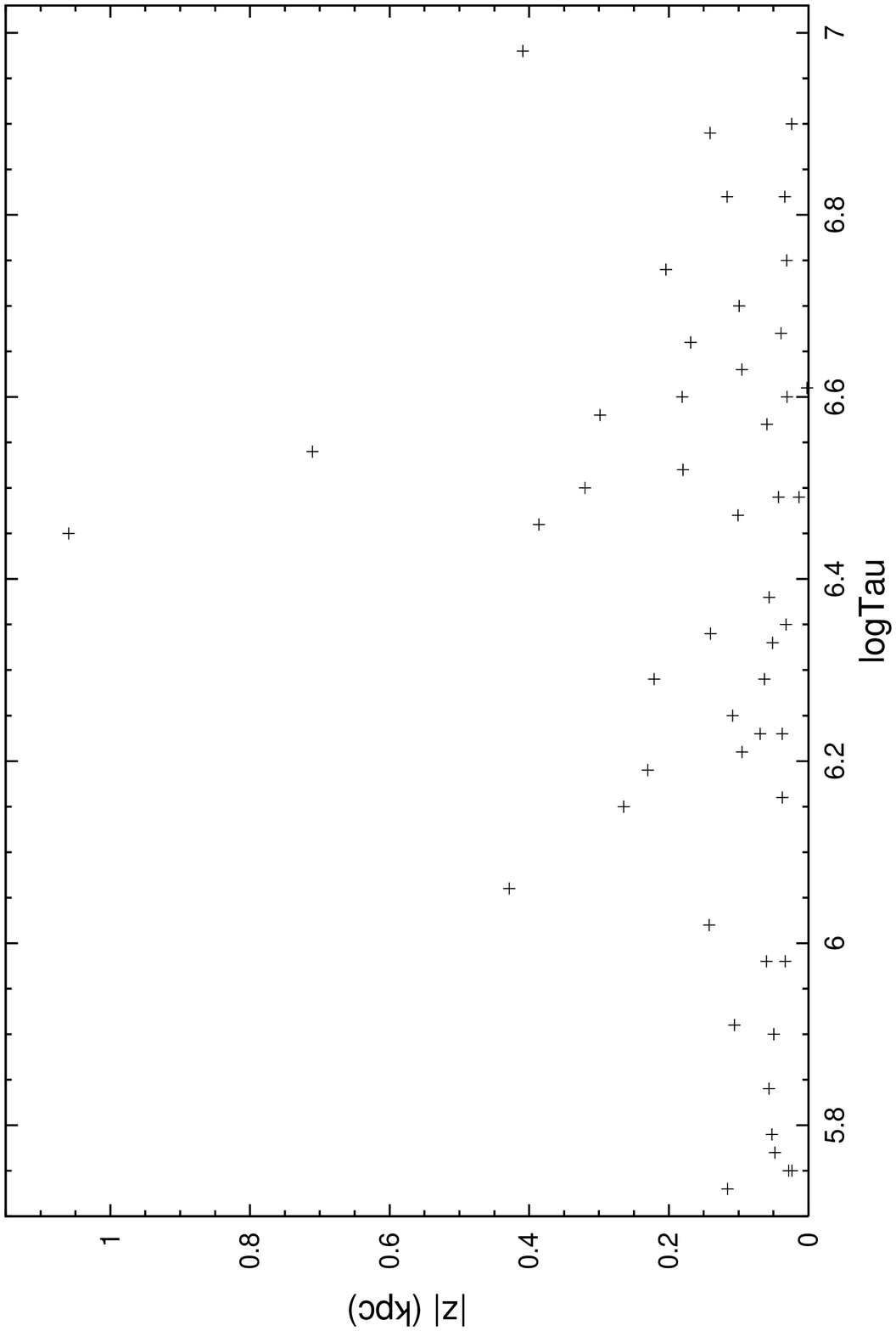,width=14.5cm,height=19cm,angle=0}
%\end{tabular}
\end{figure}

\clearpage
\begin{figure}
\psfig{file=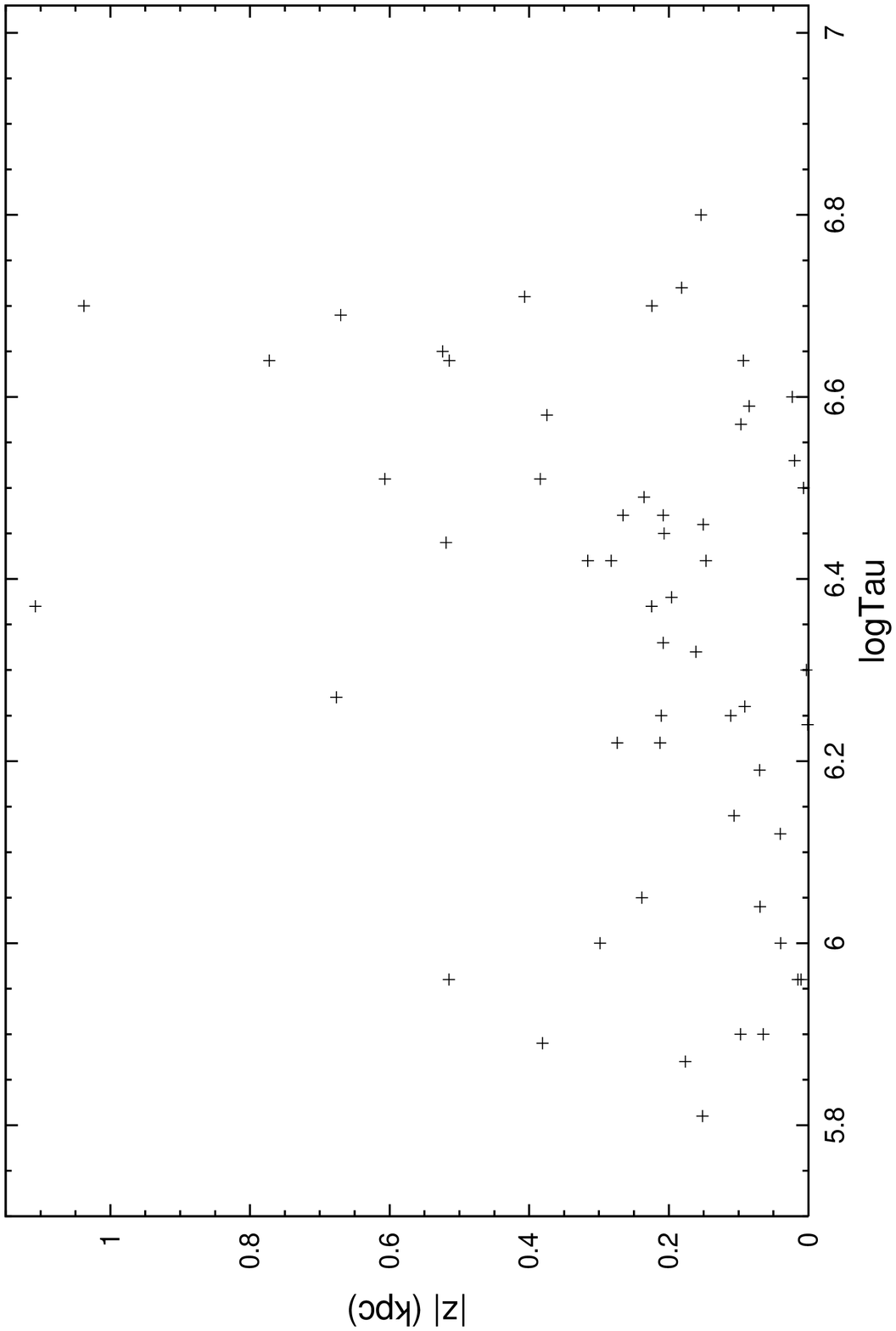,width=14.5cm,height=19cm,angle=0}
%\end{tabular}
\end{figure}

\clearpage
\begin{figure}
%\begin{tabular}{l}
\psfig{file=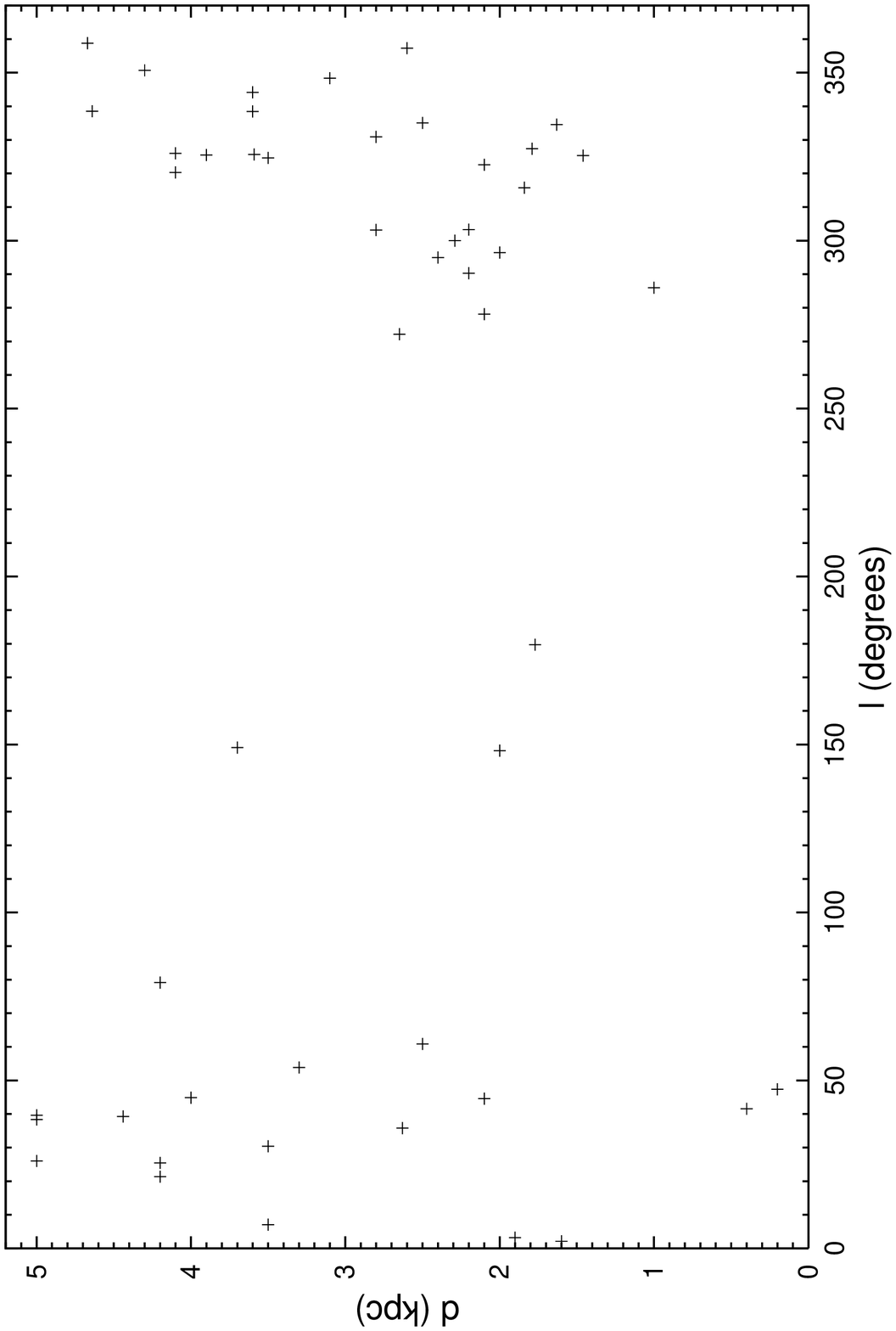,width=14.5cm,height=19cm,angle=0}
\end{figure}

\clearpage
\begin{figure}
\psfig{file=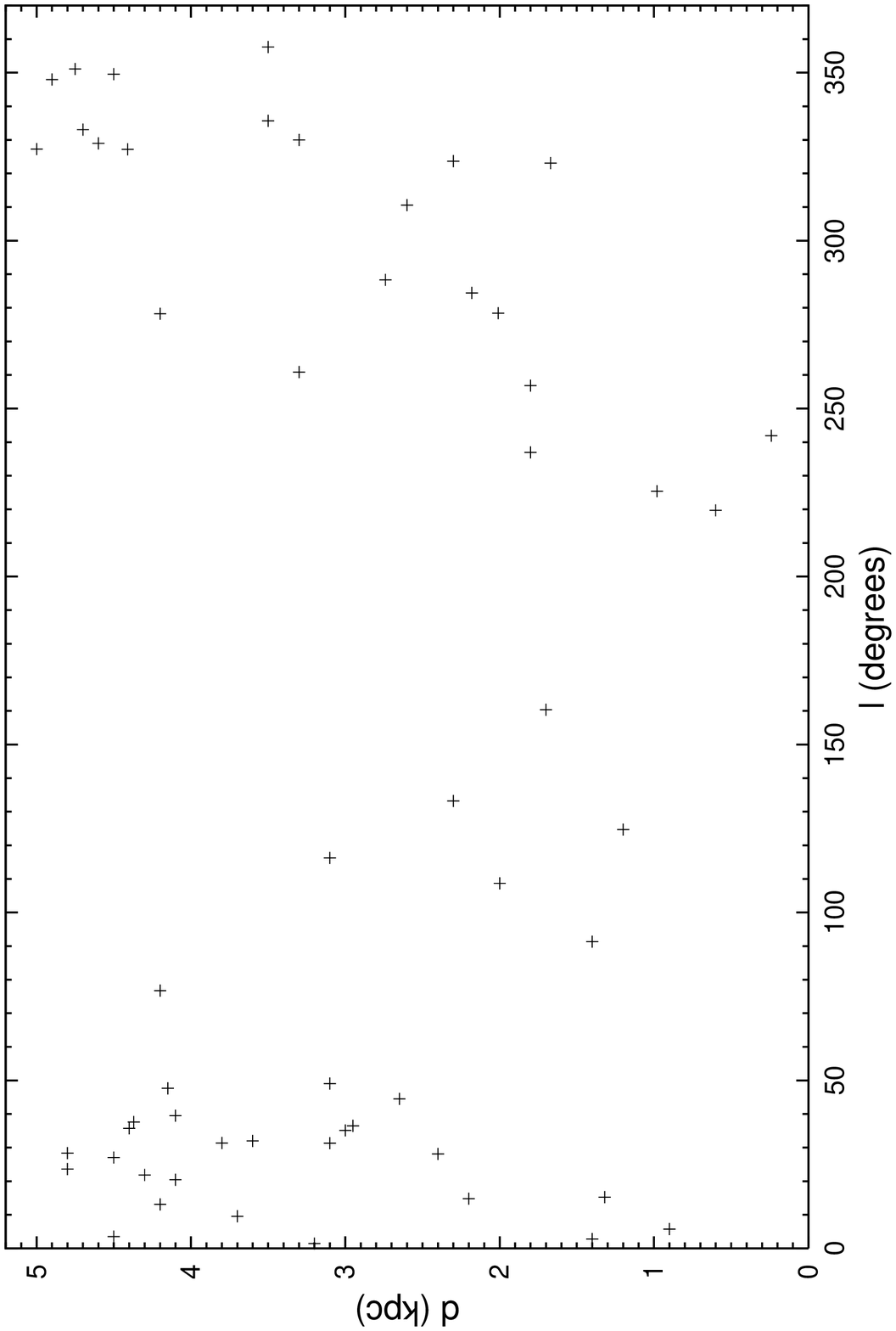,width=14.5cm,height=19cm,angle=0}
%\end{tabular}
\end{figure}

\clearpage
\begin{figure}
%\begin{tabular}{l}
\psfig{file=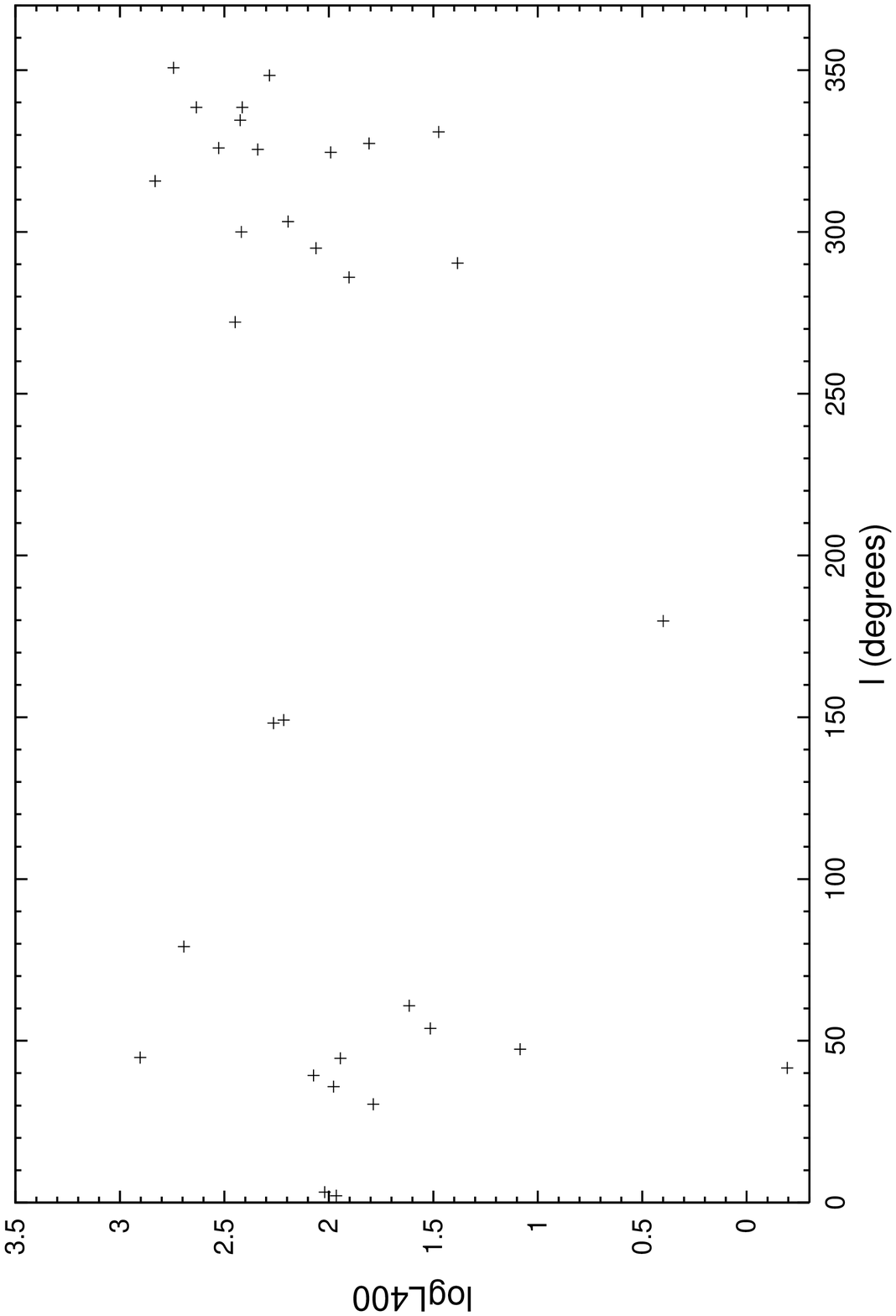,width=14.5cm,height=19cm,angle=0}
\end{figure}

\clearpage
\begin{figure}
\psfig{file=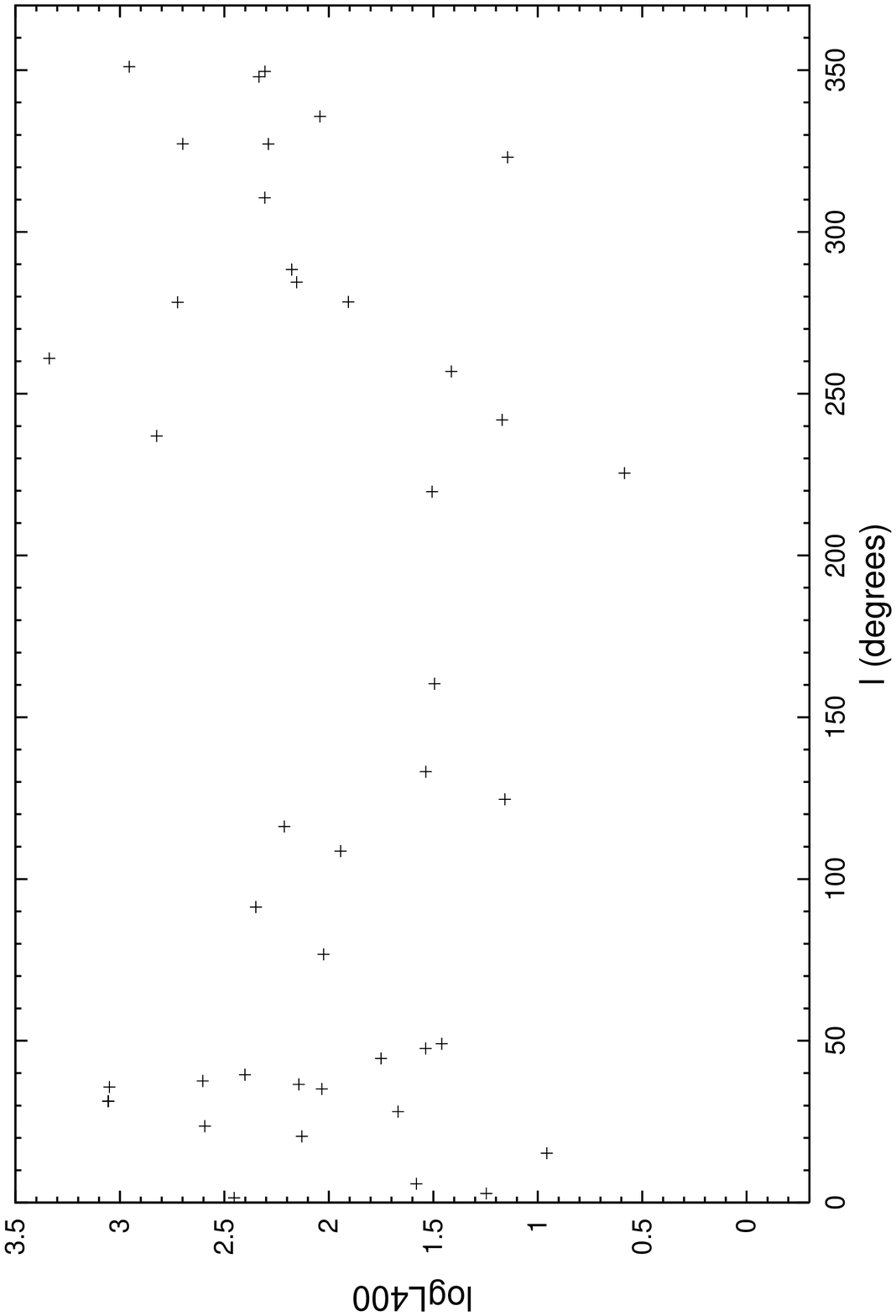,width=14.5cm,height=19cm,angle=0}
%\end{tabular}
\end{figure}

\clearpage
\begin{figure}
%\begin{tabular}{l}
\psfig{file=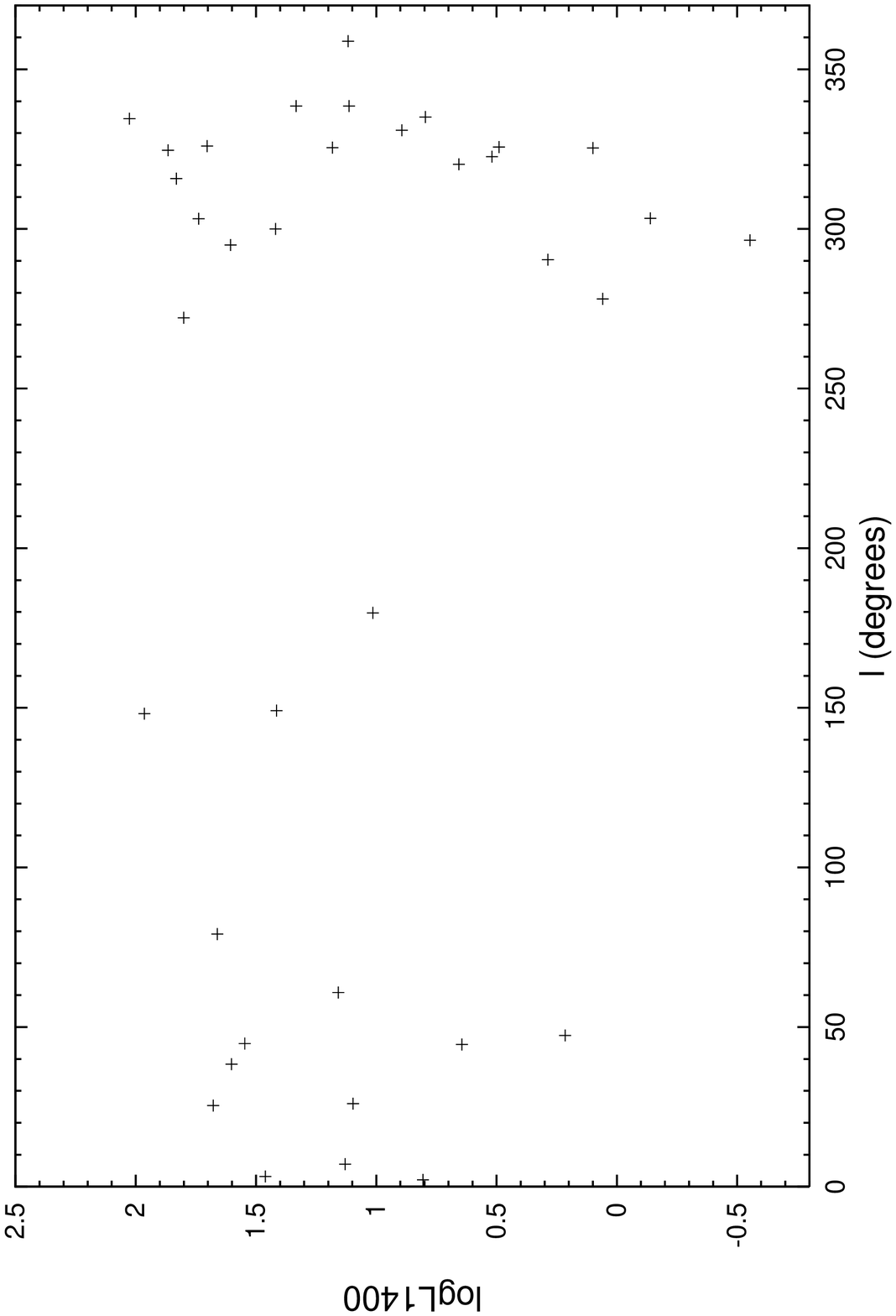,width=14.5cm,height=19cm,angle=0}
\end{figure}

\clearpage
\begin{figure}
\psfig{file=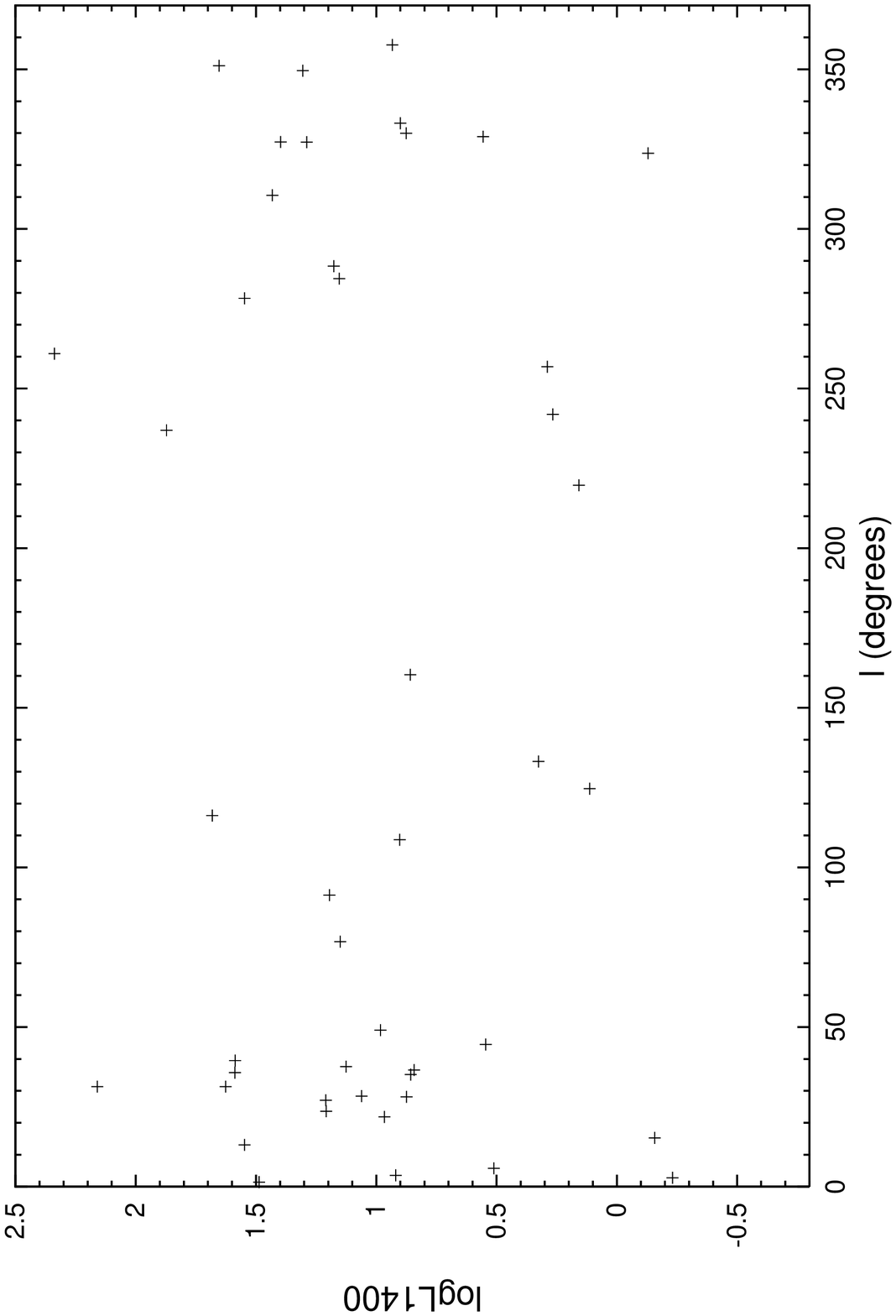,width=14.5cm,height=19cm,angle=0}
%\end{tabular}
\end{figure}

\end{document}